\documentstyle[12pt,aaspp, flushrt, psfig]{article}

\def\ls{{_<\atop^{\sim}}}

\def\cgs{ ${\rm erg~cm}^{-2}~{\rm s}^{-1}$ } 

\begin{document}
%%%%%%%%%%%%%%

\title{150~keV Emission from PKS2149$-$306  with BeppoSAX}
\author{Martin Elvis$^1$, Fabrizio Fiore$^{1,2,3}$,  
Aneta Siemiginowska$^1$, Jill Bechtold$^4$, Smita Mathur$^1$ \& 
Jonathan McDowell$^1$}

\affil {$^1$Harvard-Smithsonian Center for Astrophysics, 
Cambridge MA 02138, USA }

\affil {$^2$ Osservatorio Astronomico di Roma \\
via Osservatorio, Monteporzio-Catone (RM), I00040 Italy}

\affil {$^3$ BeppoSAX Science Data Center\\
via Corcolle 19, Roma I00100 Italy}

\affil {$^4$ Steward Observatory}

\authoremail{elvis@head-cfa.harvard.edu}
\authoremail{fiore@head-cfa.harvard.edu}

\author{\tt Astrophysical Journal 544, 545 (2000)}

\begin{abstract}
A BeppoSAX observation of the z=2.34 quasar PKS2149$-$306
produced a strong signal in the high energy PDS instrument up to
a maximum observed energy of nearly 50~keV, 150~keV in the quasar
frame. The Beppo-SAX spectrum spans almost 3 decades
(0.3$-$150~keV, quasar frame) and shows an extremely hard
($\alpha$=0.4$\pm$0.05) X-ray spectrum above 3~keV (comparable to
XJB the X-ray background slope), and either a softer
($\alpha$=1.0$^{+0.6}_{-0.3}$) low energy component, or an
ionized absorber at zero redshift.  No evidence is seen of an
Fe-K emission line (EW$<$167~eV at 6.5~keV quasar frame), a
Compton hump ($R<0.3$). A bremsstrahlung fit gives
kT(rest)=46$^{+32}_{-16}$~keV, similar to the X-ray background
value, and a high energy cut-off power law requires
E$_{cut}>$120~keV (quasar frame).

The SED of PKS~2149$-$306 shows two peaks at $\sim$10$^{12\pm
0.5}$Hz and $\sim$10$^{21\pm 1.0}$Hz ($\sim$0.3~mm and
$\sim$4~MeV), strongly resembling a low energy cutoff BL~Lac
object (LBL). The ratio of the two peaks shows an extreme Compton
dominance ($C_D=1.4\pm 0.4$), as in flat spectrum radio quasars
(FSRQs). The presence of an additional `optical/UV big bump'
component may provide photons that cool the jet, suppressing the
radio emission.

\end{abstract}

%
%%%%%%%
\section{Introduction}

High redshift quasars are of interest not only for their `record
setting' quality, but also because they can tell us about the
formation of quasars and about conditions in the universe at
early times (Turner 1991, Hamann \& Ferland 1994).  We should
also be able to learn about the quasar emission processes by
comparing the continua of quasars at low z with those of their
cousins at the epoch at which the characteristic quasar
luminosity peaks around z$\approx$2 (Boyle et al.  1987). Since
small changes in physical conditions can greatly alter the tails
of particle energy distributions the extreme high energy
continuum may be a particularly good place to look for
evolutionary effects.

So far though X-ray spectra of high redshift quasars from ROSAT
(Elvis et al., 1994a, Fiore et al., 1998a) and ASCA (Elvis et al.,
1994b, Cappi et al., 1997) have been few, and of limited
signal-to-noise. Even so they have begun to show redshift
dependent features that challenge models and require further
XJB investigation: low energy cut-offs were discovered with ROSAT
(Elvis et al., 1994a), and appear to be intrinsic (Fiore et al.,
1998a, Elvis et al., 1998) and related to ultraviolet absorbing
outflows; no Compton reflection component has been seen in z$>$1
quasars ($\Omega_d/2\pi\leq 0.4$ [90\%], Elvis et al. 1994b,
Nandra et al., 1995, Siebert et al., 1995, see also Williams et
al (1992) for low z examples); nor is a 6.4 keV (rest frame) iron
line required by the fits (EW$<$20~eV [90\%] Siebert et al.
1995; EW$<$120~eV (90\%) Elvis et al., 1994b).  These properties
differentiate high z quasars from typical low redshift Seyfert
galaxies which have strong iron K-lines (EW=100-300 eV), and a
strong Compton hump that requires half the sky (as seen from the
X-ray source) to be covered with Compton-thick `cold' material
($\Omega_d/2\pi$=1, Nandra \& Pounds 1994 and references
therein).  This suggests a different structure within high
luminosity quasars.

We observed PKS2149$-$306 as part of an AO1 GO program on
BeppoSAX in order to examine these issues. At z=2 the Compton
hump region is well within the BeppoSAX (Boella et al., 1997a)
1.3--10~keV Medium Energy Concentrator Spectrometer (MECS, Boella
et al., 1997b) band, making a measurement feasible. Moreover
the good sensitivity of the high energy Phoswich Detector System
(PDS, FWHM=1.4~degrees, Frontera et al., 1997) on BeppoSAX allows
the extension of the spectrum to energies (15--100~keV) where
a Compton hump, bremsstrahlung and a power law can be more
readily distinguished

PKS2149-306 is a z=2.34 flat spectrum
($\alpha^{4.85~GHz}_{2.7~GHz}=0.0\pm 0.14$, where $f_{\nu}\propto
\nu^{-\alpha}$, NED), core-dominated
\footnote{
The 2.3 Ghz data of Morabito et al (1986) shows the source to be
$>70 $ \% compact on a 3mas scale (24~pc at z=2.34,
$H_0$=50~km~s$^{-1}$Mpc$^{-1}$, $\Omega$=1), using the Parkes 2.7
Ghz total flux (Quiniento \& Cersosimo 1993) as a comparison.
}
radio-loud quasar with $m_V$=17.9. PKS~2149--306 is
extraordinarily bright in the ROSAT Sky Survey
(f$_X$=10$^{-11}$\cgs, Siebert et al., 1995, Brinkmann et al.,
1995b, Schartel et al., 1996).  Being bright, PKS2149$-$306 also
promised to give the best possible limits on high energy emission
from quasars. Instead, it gave us a clear detection.  We present
here the BeppoSAX 0.1-130 keV data for PKS2149--306, along with
the broad band radio to $\gamma$-ray spectral energy distribution
(SED) of the quasar, and we compare these with recent Blazar
models.

%%%%%%%
\section{Observations and data analysis}

BeppoSAX observed PKS2149$-$306 from 1997 October 31 through 1997
November 1.  The two imaging detectors - the 1.3-10 keV MECS and
the 0.1-10~keV Low Energy Concentrator System (LECS, Parmar et
al., 1997) - detected the quasar with high signal-to-noise at the
center of the their field-of-view, at 0.105~ct/s and 0.05~ct/s
respectively. (With exposures of 39,439~s and 17,862~s
respectively.)  Unexpectedly, the co-aligned Phoswich Detector
System (PDS, Frontera et al., 1997) also detected a clear signal
in its 15-100~keV bandpass of 0.26$\pm0.05$~ct/s (in an effective
on-source exposure time of 16,676~s).

Standard data reduction was performed using the SAXDAS software
package (1997).  In particular, data are linearized and cleaned
from Earth occultation periods and unwanted periods of high
particle background (due to satellite passages through the South
Atlantic Anomaly, SAA).  Data from MECS units 2 and 3 were
combined after gain equalization (unit 1 has been inoperative
since 1997 May 7).  The MECS and LECS pulse height spectra of
PKS2149$-$306 were extracted using 3~arcmin and 8~arcmin radius
extraction regions, respectively. These radii maximize the
signal-to-noise ratio below 1 keV in the LECS and above 2 keV in
the MECS. Background was taken from high Galactic latitude blank
fields using the same extraction regions as for the source
regions. Although background is negligible for this source, we
checked that the level of the background in regions surrounding
the source is within 10\% of that recorded in the same regions in
the blank fields.

For the PDS the problem of accurate background subtraction is
crucial since the PDS is a non-imaging detector. The PDS uses
rocking collimators, one for each of two pairs of detectors, that
switch positions quickly every 92~s so that one of the pairs is
always pointed at the source, while the other is monitoring the
background.  The spectra from the four PDS crystals were summed
together. The net source spectrum was then obtained by
subtracting the `off-source' background from the `on-source'
counts, scaling by the exposure time. The background in the
equatorial BeppoSAX orbit varies by only $\sim$10-20\%, thanks to
shielding by the Earth's particle belts (this was the main reason
for selecting an equatorial orbit for BeppoSAX), and has an
active anti-coincidence CsI(Na) particle shield (which is also
used to detect gamma-ray bursts, Frontera et al., 1997). The main
background variations are recorded just after the passages
through the SAA. (The instrument is switched off while in the
SAA). The first five minutes after SAA passages are excluded from
the data analysis to allow an automatic gain control system to
return the high voltages to their normal working value.  This
system keeps the pulse-height to energy scale constant to 1\%.
Using the rocking technique it is possible to take into account
accurately the residual background variations.  The magnitude of
residual systematic errors in PDS spectra have been evaluated by
Guainazzi and Matteuzzi (1997) at $\ls 0.05$ counts s$^{-1}$,
using deep observations of blank fields. Even if this maximum
residual is subracted from the observed count rate, the signal to
noise in this observation would remain $>4\sigma$.  The source is
detected at $>3 \sigma$ up to 40-50 keV, which is about 150 keV
in the quasar rest frame.

The chance of finding a source in any given 2~sq.degree, the FWHM
beam area of the PDS, is small. The HEAO-1 A4 all sky catalog
(Levine et al., 1984) lists just 7 high Galactic latitude
($|b|>20^{\circ}$) sources in the 13-80~keV band, which is
closely comparable to the PDS band, down to a flux of
2$\times$10$^{-10}$erg~cm$^{-2}$~s$^{-1}$ (10mCrab). The
PKS2149$-$306 signal is 7.5 times fainter so, assuming a
logN-logS slope of $-$1.5, we expect a chance coincidence rate of
1.4\%, hence we believe the PDS signal does come from
PKS2149--306. Our confidence in the identification of the PDS
signal with PKS2149$-$306 is increased further by the agreement
to better than 10\% of the normalizations of the PDS and MECS
spectra and the similarity of their slopes (see next section).
The EGRET source reported later is too far away to produce the
PDS signal (see \S 4).

%%%%%%%
\section{X-ray Spectral Fitting}
\label{fits}

Spectral fits were performed using the XSPEC~10.0 software
package and versions of the response matrices made public on 1997
August 31.  PI channels in MECS and LECS spectra are rebinned
sampling the instrument resolution with the same number of
channels at all energies and ensuring at least 20 counts per bin
for the other.  This guarantees the applicability of the $\chi^2$
method in determining the best fit parameters, since the
distribution in each channel can be considered Gaussian.
 
Constant factors have been introduced in the model fitting in
order to take into account the intercalibration systematics
between instruments ({\em BeppoSAX Cookbook}, Fiore et al. 1998b).
%In the fits we used the two MECS as reference instruments, and
%constrained the LECS and PDS normalizations to lie in the ranges
%0.7-1.0 and 0.80-0.95 relative to the MECS, respectively [{\sc
%Fiore correction}].  
If we use the two MECS as reference instruments, then
the LECS and PDS normalizations lie in the ranges
0.7-1.0 and 0.80-0.95 relative to the MECS, respectively.

The (observed frame) energy ranges used for the fits are: 0.1-4
keV for the LECS, 1.65-10.5 keV for the MECS, and 14-130 keV for
the PDS. All the results given in table~1 are in the observed
frame. Quasar frame values are given in the text where
appropriate.  Errors in Table 1 are 90 \% confidence intervals
for 1 interesting parameter.  A simple power-law plus absorption
model gives a reasonably good fit (Figure~1, Table~1), but with 
systematic deviations at low energies.

PKS2149--306 appears to have faded slightly over the three years
between the obsevations: the 2-10 keV BeppoSAX flux, using the
simple power law fit, is ($8.0\pm0.2)\times10^{-12}$\cgs, 80\% of
the flux recorded in the 1994 October ASCA observation
($9.9\times10^{-12}$\cgs, Cappi et al 1997); and the
monochromatic 2 keV BeppoSAX flux is ($2.8\pm0.1)\times
10^{-12}$\cgs~keV$^{-1}$, 76\% of the 1990/1991 ROSAT RASS
detection at ($3.7\pm0.3)\times 10^{-12}$\cgs~keV$^{-1}$
(Schartel et al. 1996).

The LECS shows a clear excess of counts below about 1~keV. This
soft excess seems in contradiction with the above Galactic N$_H$
reported by Yaqoob et al. (1999). While variability in the
11~months (quasar frame) between the two observations is
possible, we also note that the low energy calibration of ASCA 
is uncertain
\footnote{see:
http://heasarc.gsfc.nasa.gov/docs/asca/watchout.html
\newline
http://heasarc.gsfc.nasa.gov/docs/asca/cal\_probs.html
}
, and we have more confidence in the Beppo-SAX value (Parmar et
al., 1997, Mineo et al., 2000). The presence of a low energy
excess means that the X-ray spectrum cannot give a reliable
Galactic absorption value, so we fix the absorbing column to the
Galactic 21~cm value (2.2$\times$10$^{20}$cm$^{-2}$, Dickey \&
Lockman 1990). The resulting energy spectral index,
$\alpha=0.40\pm0.04$, is extremely flat (c.f. Sakano et al. 1998,
Sambruna et al., 1999), although consistent with other high
redshift radio-loud quasars measured with ASCA from 2-10~keV ,
including PKS2149$-$306 itself ($\alpha$=0.42$^{+0.03}_{-0.02}$
Cappi et al. 1997, 0.54$\pm$0.05 Yaqoob et al. 1999). This slope
is close to that of the X-ray background (Marshall et al., 1980)
although this class of quasar is too rare to integrate to the
X-ray background (see Elvis et al., 1994 and
references. therein).

%%%%%%
\begin{figure*}
\centerline{ 
\psfig{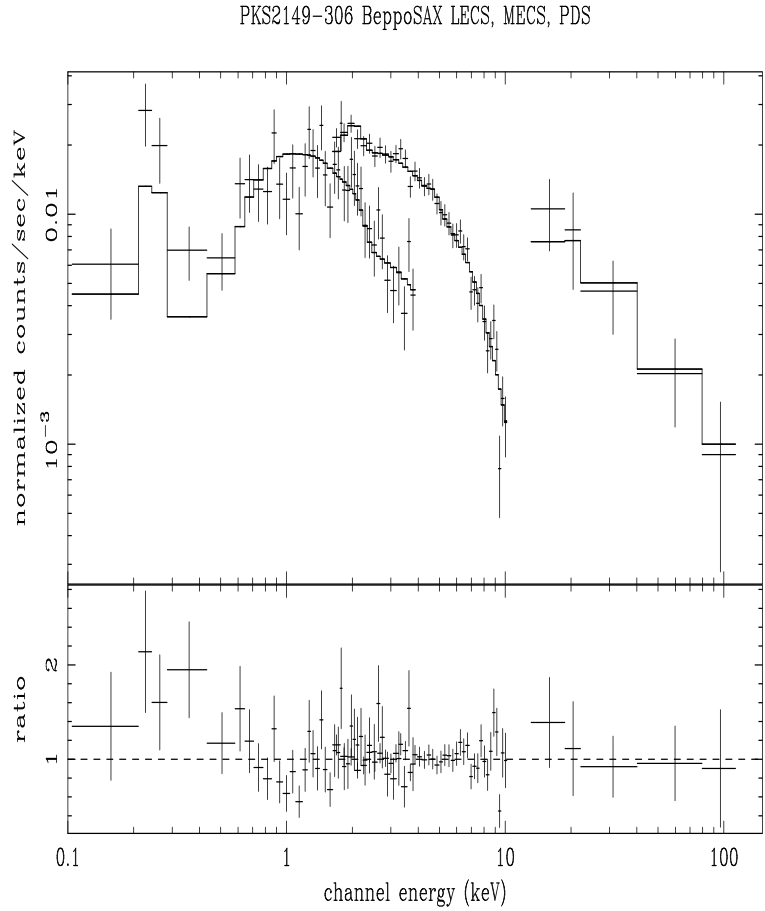}
}
\caption{The 0.1-130~keV (observed frame) BeppoSAX LECS, MECS and
PDS spectrum of PKS2149-306 fitted with a single power law model
reduced at low energy by Galactic absorption. The lower panel
shows the ratio between data and model.}
\label{fig1 }
\end{figure*}
%%%%%%

A better fit can be had with a broken power-law to account for
the slight upturn in the LECS spectrum below $\sim$1~keV (Table
1).  In this case the steeper low energy slope is
$2.0^{+2.0}_{-1.0}$ allowing the absorbing column to be free to
vary, or $1.0^{+0.6}_{-0.3}$ fixing $N_H$ to its Galactic
value. Both slopes are consistent with the ROSAT RASS value of
$1.24\pm0.80$ (Schartel et al., 1996).  The break energy is about
0.8 keV (observed frame, 2.7~keV quasar frame) in both cases. The
high energy index is indistinguishable from the single power law
index.

An apparent soft excess can also be generated though an
absorption edge.  In the quasar frame there is no sign of an
ionized oxygen absorber, although at $z=2.34$ the energy of the
deepest edges usually seen in low luminosity Seyfert galaxies
(OVII and OVIII) fall at 0.22 and 0.26 keV (observer frame)
respectively, close to the carbon edge (0.283 keV) produced by
the LECS window, and are therefore affected by large systematic
errors.  A fit with an absorption edge and a single power law
(Table~1) gives a good $\chi^2$. The edge energy is 2.8~keV in
the quasar frame, which has no ready interpretation, but the
observed energy of 0.8~keV is consistent with either OVIII at
$z<0.2$ or OVII at $z=0$. The optical depth is large,
$\tau\sim1$, implying a total hydrogen column of a few $10^{22}$
cm$^{-2}$ assuming cosmic metal abundances. This is about one
hundred times what it is seen from the Galactic 21 cm line.  The
high energy slope in this fit is consistent with the value from
the broken power law fit.

No line is detected around 2~keV, the redshifted iron K$\alpha$
energy.  The 90 \% upper limits to the emission line equivalent
width are 50 eV and 25 eV (observer's frame; 167~eV and 84~eV
quasar frame) for the 6.4 and 6.7 keV (quasar frame energies)
iron K$\alpha$ lines respectively. No line is seen at 5.1~keV
(observer's frame, 17~keV quasar frame) with a limit of 63~eV
(90\% confidence, observer's frame), or 210~eV (quasar
frame). This compares with the 298$^{+202}_{-205}$~eV (quasar
frame, 90\% errors for two interesting parameters) measurement
reported at thsi unusual energy by Yaqoob et al. (1999) from 1994
October ASCA data.

The presence of a steeper low energy component or absorption edge
has little effect on the high energy slope.  Since this steeper
component is only important in the LECS data we shall fit
only the MECS and PDS data when investigating the high energy
component, as we do in the rest of this section. This has the
advantage of removing extra free parameters from the models,
which usefully decreases the allowed parameter space.

To determine the overall 5-150~keV rest frame spectral shape we
next fitted three XSPEC models: a power-law with an exponential
cutoff ({\tt cutoffpl}); a Compton reflection model for neutral
reflectors ({\tt pexrav}, Magdziarz \& Zdziarski 1995); and a
thermal bremsstrahlung model.  In {\tt pexrav} the redshift was
frozen at the quasar redshift, the disk inclination was frozen at
cos~$i$=0.45, and the cosmic abundances of Anders \& Ebihara
(1982) were assumed. For both models the local absorption was
fixed at the Galactic value.

The Compton reflection model (Table 1) gave a 90\% upper limit of
0.3 to the relative normalization, $R$, of the reflected
component
\footnote{$R = \Omega/2\pi$, for an isotropic source above a flat
infinite disk}
.  This compares with typical values of 1.0 in Seyfert galaxies
(Matt 1998).  The cut-off power-law model (Table 1) requires the
cut-off energy, E$_c$, to be $>$36~keV (90\% confidence,
observed frame), or $>$120~keV in the quasar frame.  A
bremsstrahlung fit gives a rest frame temperature of
46$^{+32}_{-16}$~keV (90\% confidence, 1 interesting parameter),
similar to that found in several z=3 radio-loud quasars in the
2-10~keV band with ASCA (Cappi et al., 1998). The Beppo-SAX
result suggests that these temperatures are not merely an
artifact of the upper energy limit of ASCA. As for ASCA, a
bremsstrahlung model is a slightly worse fit than power-law
models.

%%%%%%
\begin{table*}[ht]
\caption{\bf Spectral fits (observed frame)}
\begin{tabular}{lccccc}
\hline
model & $N_H^a$ & $\alpha_H$ & $\alpha_S$ & additional 
& $\chi^2$(dof) \\ 
&&&&parameters&\\
\hline
\multicolumn{6}{l}{\em LECS, MECS, PDS fits} \\
Power law & 2.2 FIXED & 0.4$\pm$0.04 & -- & -- & 92.3(91) \\
Broken p.l. & 5.0$^{+5.0}_{-2.8}$ & 0.39$\pm$0.05 & 2.0$^{+2.0}_{-1.0}$ &
$E_b=0.8\pm$0.4 & 81.2(88)\\
Broken p.l. & 2.2 FIXED & 0.37$\pm$0.04 & 1.0$^{+0.6}_{-0.3}$ &
$E_b=0.8\pm$0.4 & 84.1(89)\\
P. L. + edge &  2.2 FIXED & 0.49$\pm0.08$& -- & $E_e=0.86\pm0.12$
& 80.9(89)\\
&&&& $\tau=0.9\pm0.5$ & \\ 
\hline
\multicolumn{6}{l}{\em MECS, PDS fits} \\
Power law & 2.2 FIXED & 0.41$\pm$0.05 & -- & -- & 46.9(51) \\
Cut-off p.l. & 2.2 FIXED & 0.41$\pm$0.05 & -- & $E_c>36$ & 46.9(50) \\
Compton refl. &  2.2 FIXED & 0.41$\pm$0.05 & -- & $R<0.3$ & 46.9(50) \\
Bremsstrahlung & 2.2 FIXED & T$^b$=46$^{+32}_{-16}$ & -- & -- & 54.9(51) \\
\hline
\end{tabular}

$E_b$=break energy; $E_c$=cut-off energy; $E_e$=edge energy;
$\tau$=edge optical depth.  
$^a$ in $10^{20}$ cm$^{-2}$; 
$^b$ in keV (quasar frame).
\end{table*}

%%%%%%%
\section{The Spectral Energy Distribution of PKS2149$-$306}

The addition of the broad band BeppoSAX spectrum from 0.1-50~keV
(observed) allows the construction of a well-sampled Spectral
Energy Distribution (SED) since PKS2149$-$306 is already
XJB well-observed at radio wavelengths, as shown by a NED
\footnote{The NASA/IPAC Extragalactic Database (NED) is operated
by the Jet Propulsion Laboratory, California Institute of
Technology, under contract with the National Aeronautics and
Space Administration.}
search.  Additional radio (Quiniento et al., 1993) and millimeter
data (Steppe et al., 1988) fill out the long wavelength SED.
Although PKS2149--306 has a flat radio spectrum, 
Cersosimo et al (1994) include PKS2149-306 as a candidate
GigaHertz Peaked Spectrum (GPS) radio source, because of a mild
curvature in its radio spectrum. PKS2149--306 also has low radio
polarization ($<1.7$\% at 1.4~MHz, Condon et al 1998).  These
two characteristics of GPS sources make it unlikely that such
sources are dominated by beamed emission (O'Dea 1998).
PKS~2149--306 though, has a Blazar-like SED and is unlikely to be
a normal GPS source. The curvature in the PKS2149--306 radio
spectrum may instead be due to the Compton downshifting of the
originating electron spectrum, caused by a quenching UV source
(see \S\ref{dominant})

In the 100$\mu$m to 0.1$\mu$m range there are only IRAS Faint
Source Survey (FSS, Moshir et al., 1989) upper limits and an
optical spectrum (Wilkes et al., 1983, Wilkes 1997, private
communication).  These few data points though are importantly
constraining.  A search of the IRAS FSS noise plates gives
3~$\sigma$ upper limits of 130~mJy, 96~mJy, 70~mJy and 600~mJy at
12$\mu$m, 25$\mu$m, 60$\mu$m and 100$\mu$m respectively. These
limits are stable against small changes in position, so that a
complex background (e.g. from Galactic cirrus) is not causing
spurious values.

The Compton Gamma Ray Observatory EGRET summed phase 1-4
(1991--1995) exposure (9$\times$10$^8$~seconds) of the field shows
no evidence for emission from PKS2149--306. A point source fit at
the position of the quasar gives a 3 sigma upper limit of
$6.9\times 10^{-8}$ photons~cm$^{-2}$s$^{-1}$ (30~MeV-3~GeV). We
do recover the well known BL Lac object PKS 2155--304 (3EG
J2158--3023, Hartman et al 2000) with a flux of ($1.2\pm
0.3)\times 10^{-7}$ photons~cm$^{-2}$s$^{-1}$
(30~MeV-3~GeV). However, this source lies 1.7 deg from
PKS2149--306, well outside the 99~percent confidence region of
the EGRET source position. The PDS reaches a transmission of
50~percent at 0.7 degrees and is effectively at zero transmission
by 1.7 degrees, so PKS 2155--304 cannot be the origin of the PDS
signal.

The complete SED of PKS2149$-$306 (Figure 2) shows how unusually
hard and energetically dominant the hard X-ray, PDS, spectrum is
when compared with normal radio-loud quasar SEDs (dot-dash curve,
Elvis et al., 1994c). The whole Beppo-SAX spectrum lies an order
of magnitude above the normal level for radio-loud quasars.  The
X-ray `soft excess' (E$<$3~keV in the quasar frame) {\em appears}
to be a smooth continuation of the $\alpha\sim1.0$ optical slope
(with the caveat that an ionized absorber could fake an excess,
see \S 3).  In contrast the 3-150~keV slope is harder by
$\Delta\alpha\sim0.5$.  However, unless the source varies by a
factor greater than 50 at $\sim$1~GeV, this hard X-ray MECS/PDS
slope cannot extend up to EGRET energies, and must instead turn
down somewhere between 120~keV (the quasar frame lower limit on a
cut-off in the PDS spectrum) and 40~MeV.

%%%%%%
\begin{figure*}
\centerline{ 
\psfig{figure=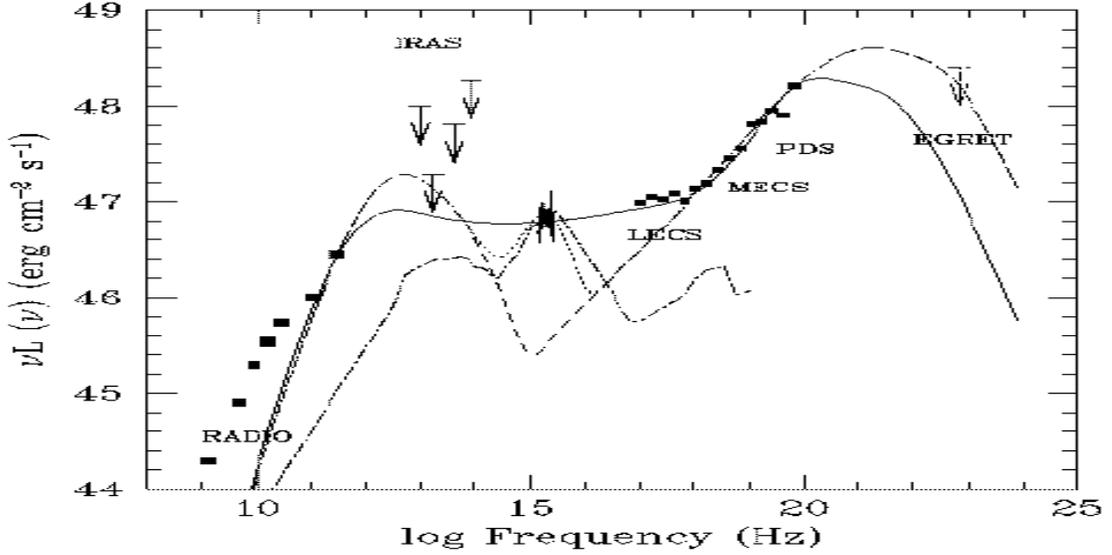,height=12cm,width=18cm,angle=0}
}
\caption{The radio to $\gamma$-ray SED of PKS2149--306. The solid
line is a spline fit close to one of the synchro-compton models
from Ghisellini et al.(1998). The long dashed line is a spline
interpolation through the highest frequency radio point and the
X-ray data that does not exceed the IRAS and EGRET upper
limits. The two spline curves (solid and long-dash) illustrate
the uncertainty in defining $C_D$ ($=log(L_{IC}/L_{synch}$). The
thin dot-dash line is the mean SED of radio-loud quasars from
Elvis et al. (1994c), normalized to the optical, to show how
PKS2149--306 is a factor $\sim$10 more X-ray loud than is
typical.  The short-dash line shows an illustrative accretion
disk model (for $M_{bh}=10^{10}M_{\odot}$,
$\dot{m}/\dot{m_{Edd}}$=0.1) added to the long-dash curve,
showing the likely large UV bump in PKS2149--306.}
\label{fig2 }
\end{figure*}
%%%%%%

The SED of PKS~2149--306 (figure~2) seems to have two peaks at
about $10^{12}$ and $10^{21}$ Hz.  This `two peaks' SED shape is
similar to those of blazars (Fossati et al. 1998), strongly
suggesting that the observed SED, outside the optical band, is
dominated by a blazar continuum originating in a jet.  The solid
and dashed lines on the SED (figure 2) are spline interpolations
of the highest frequency radio point and the X-ray data above 2
keV constrained not to exceed the IRAS and EGRET upper limits.
The solid line is a spline fit close to one of the
synchro-Compton models from Ghisellini et al.(1998). The lines
are chosen to indicate the uncertainty in determining the ratio
of the two peaks.  (The lines are not fits to the data with a
physical model, and are only useful to guide the eye.)  The two
spline curves define the ranges of the two peak frequencies and
amplitudes. The $\gamma$-ray peak lies at a few MeV,
$\nu_{peak}(IC)=$10$^{21.0\pm 1.0}$Hz, given the EGRET upper
limit.  The peak of the low frequency, radio-mm, peak is in the
sub-millimeter range, $\nu_{peak}(synch)=10^{12.0\pm0.5}$
Hz. These are both unusually low frequencies. The ratio of the
two peak luminosities, the Compton dominance, $C_D$ (=$log
L_{IC}/L_{synch}$, Ghisellini et al., 1998 )$= 1.4\pm0.4$, is
near the top of the known range.

%%%%%%%%%%%%%%%%%%%%%%%%%%%%%%%%%%%%%%%%%%%%%%%%%%%%%%%
\section{Discussion}

%%%%%%%%%%%%%%%%%%%%%%%
\subsection{PKS~2149--306: An Extreme Compton Dominant Quasar}
\label{dominant}

The detection of hard X-ray emission in PKS~2149$-$306 reaches an
energy some 4 times higher than detected in any other z$>$1
quasar, except for the EGRET blazars. Moreover the X-ray to
optical ratio of PKS~2149--306 is large, a factor 10 larger than
normal for radio-loud quasars. Expressed as $\alpha_{OX}$, the
optical (2500\AA ) to X-ray (2~keV) slope, 0.81, is unusually
small. Only two radio-loud quasars out of some 250 in the survey
by Brinkmann et al. (1995a) have $\alpha_{OX}<0.9$.  Clearly
there is something unusual about this quasar.  The SED of
PKS~2149$-$306 is not that of a typical, unbeamed, AGN. The `two
peaks' SED shape has not been seen in `normal' unbeamed quasars
(Elvis et al., 1994, Mattox 1994), although existing limits are
surprisingly weak), but is common in relativistically beamed
objects, both BL~Lacs and quasars (Padovani \& Giommi 1996).  The
absence of reflection components (Compton hump and iron line)
strongly limits the contribution to the observed flux of any
Seyfert-like component, although these features are also weak in
radio-quiet high luminosity quasars (Iwasawa \& Taniguchi 1993).

The `two peaks' SEDs of BL~Lac objects have often been considered
within the Synchrotron-Inverse Compton formalism (e.g. Sambruna,
Maraschi \& Urry 1996). In this scheme the low frequency
(radio-mm) peak is the primary synchrotron peak, while the high
frequency ($\gamma$-ray) peak comes from photons Inverse Compton
scattered off the relativistic electrons. These scattered photons
may be the synchrotron photons themselves (the self-Compton
case), or from an external radiation source.

The frequencies at which the two components peak in $log \nu
f_{\nu} ~vs.~ log \nu$ space varies by 5 decades, which changes
their observational appearance greatly.  Fossati et al. (1998)
and Ghisellini et al. (1998) have suggested that the Blazars form
a single, smooth sequence with bolometric luminosity, from the
lower luminosity High energy cut-off BL Lacs (HBL) to the Low
energy cut-off BL Lacs (LBL) and then to the high luminosity flat
spectrum radio quasars (FSRQs, Sambruna, Maraschi \& Urry 1996,
Padovani, Giommi \& Fiore 1997). Unlike HBLs and LBLs, FSRQs show
normal quasar emission in the optical/UV, i.e. both the broad
emission lines and the `big bump' blue continuum (with a presumed
accretion disk origin) are seen. This optical emission can
occasionally dominate, as it does in PKS2149--306.  We conclude
that PKS2149--306 is a high luminosity FSRQ.

In PKS2149--306 both SED peaks lie at the extreme lower ends of
the distributions in the Fossati et al (1998) and Ghisellini et
al (1998) samples. In the Fossati et al. scheme the `Compton
dominance' (the ratio between the luminosity in the two peaks) is
inversely proportional to the frequency of the synchrotron peak,
roughly to the one half power (Figure~3). PKS2149--306 fits this
correlation, in that the large Compton dominance of PKS2149--306
agrees with low frequency position of the synchrotron peak of its
SED.

%%%%%%
\begin{figure*}
\centerline{ 
\psfig{figure=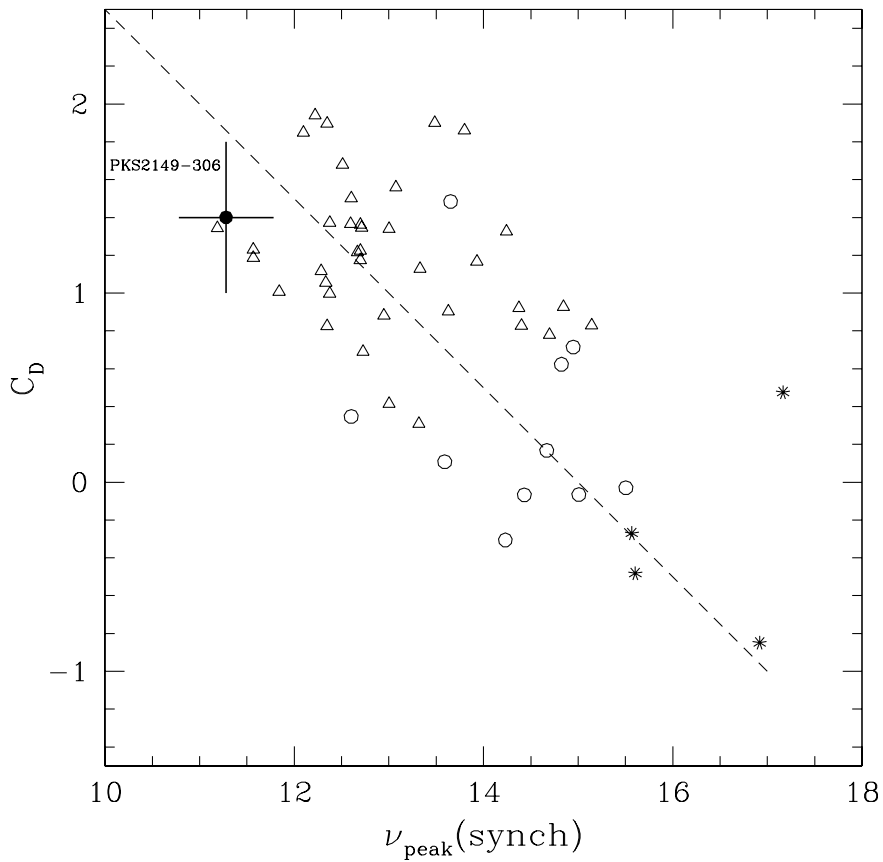,height=12cm,width=12cm,angle=0}
}
\caption{Compton dominance
($C_D=log(L_{IC}/L_{synch}$) vs. observed synchrotron peak
XJB frequency ($\nu_{peak}(synch)$, after Ghisellini et al.,
1998). PKS2149-306 is the solid point with the error bars. A
$C_D \propto \nu_{peak}(synch)^{-0.5}$ relation is shown.
Stars are HBLs (High energy cutoff BL~Lacs); open circles are
LBLs  (Low energy cutoff BL~Lacs); triangles are FSRQs (Flat
Specrum Radio Quasars).
}
\label{fig3}
\end{figure*}
%%%%%%

Ghisellini et al. go on to suggest that a strong external
radiation source, such as an accretion disk powered `optical/UV
big bump', provides a bath of cool photons that can efficiently
drain energy from jet electrons by Compton scattering them to
high energies.  The low energy photons may come directly from the
disk below (Dermer \& Schlickeiser 1993) or from above or from
the side (Siemiginowska \& Elvis 1994, Sikora et al., 1996). This
cooling makes inverse Compton losses dominate over synchrotron
losses making the radio spectrum weak relative to the X-ray, so
creating a `Compton dominant' SED.  PKS2149--306 seems to have a
large `optical/UV big bump', as expected in this model. Such
large optical/UV big bumps are unusual, but not unprecedented in
FSRQ (e.g. 3C273, Elvis et al., 1994c). An illustrative accretion
disk spectrum, that roughly fits the big bump, is shown in
figure~2 ($M=10^{10}~M_{\odot}$, $\dot{m}/\dot{m}_{Edd}$=0.1,
from the tabulation given in Siemiginowska et al. 1995).

%%%%%%%%%%%%%%%%%%%%%%%
\subsection{Low Energy Excess or Ionized Absorber?}

The soft X-ray excess in PKS~2149--306 appears unusually strong
(Figure~2). We noted in \S\ref{fits} that an absorption edge
imprinted onto a single power law gives an equally good fit.
Since ionized absorbers of the kind that could produce such an
edge are common in AGN and blazars we must consider this
interpretation on an equal footing with a `soft excess'
component. 

An absorption edge can be more tightly constrained than a soft
excess. A local z$\sim$0 absorber is unlikely since no molecular
cloud is nearby (Marscher 1988, Turner et al., 1995), nor is any
low z galaxy seen on the STScI digitization of the sky surveys.
An absorber associated with the quasar is more likely, X-ray
absorption with a similar column density is common in radio-loud
high redshift quasars (Elvis et al., 1994a, Cappi et al., 1997,
Fiore et al., 1998a, Elvis et al., 1998), and there is evidence
for high ionization oxygen absorbers of similar column density in
some low redshift blazars (Canizares \& Kruper 1984, Koenigl et
al. 1995, Sambruna et al. 1997, Grandi et al. 1997, Cagnoni \&
Fruscione 2000), though these absorbers are not always seen
(Chiappetti et al. 1999). {\em Chandra} high resolution grating
spectra however do show comparable features (Fruscione et
al. 2000, in preparation).  To produce an apparent $z$=0 absorber
would require a coincidence of outflow velocity and cosmological
redshift. A substantial ejection velocity, $v=0.6c$, is implied.
In PKS~2155--304 the absorber has a suggested velocity of $\Delta
v\sim 0.1c$ relative to the core (Canizares \& Kruper 1984) so a
high ejection velocity in PKS~2149--306 is not inconceivable.
$v=0.6c$ is larger than the $v=0.1-0.2c$ seen in broad absorption
line quasars (Turnshek 1988); it is much slower, though, than the
velocities needed to explain superluminal radio sources (Zenzus
1997), although these may be pattern speeds rather than true
expansion velocities.  Only X-ray spectra with significantly
better resolution can definitively identify absorption
edges. {\em Chandra} grating observations are scheduled.

%%%%%%%
\section{Conclusions}

We have detected the z=2.34 quasar PKS2149$-$306 up to 150~keV in
its rest frame, using the PDS on BeppoSAX. This is some 4 times
higher in energy than any other z$>$1 quasar, except for the
EGRET blazars. PKS~2149-306 is also an order of magnitude X-ray
loud relative to the optical ($\alpha_{OX}$=0.81). The $>$3~keV
spectrum of PKS~2149--306, as for several other high redshift
radio-loud quasars, is extremely hard ($\alpha=0.41\pm0.05$,
kT=46$^{32}_{16}$~keV, observed frame), comparable to the X-ray
background spectrum.

The PKS2149$-$306 SED seems to show the two peaks typical of
BL~Lacertae objects, suggesting that the X-ray 1-100 keV observed
emission is dominated by an inverse Compton component. Unusually
the SED is strongly dominated by the $\gamma$-ray peak.  A strong
`big bump' in the optical/ultraviolet seems to be present in
PKS2149--306. Possibly, as suggested by Ghisellini et al. (1998),
this bump is the source of the photons that cool the jet,
suppressing the radio emission and creating the Compton dominated
spectrum.  This makes PKS2149$-$306 an extreme example of the
luminosity based one-parameter unification scheme of Ghisellini
et al..

The low energy X-ray spectrum of PKS~2149--306 shows no
absorption above the Galactic value, contrary to the ASCA 1994
data (Yaqoob et al. 1999), showing instead a soft excess. An
absorption edge fits the data as well as a broken power-law. One
interpretation of an absorption edge is an oxygen absorber with
N$_H\sim$10$^{22}$cm$^{-2}$, blueshifted from the quasar redshift
and implying an ejection velocity, $v\sim 0.6c$. Such an absorber
would provide a link with absorbers in other high redshift
quasars, and with the absorbers in some low z blazars. However,
the data are ambiguous and higher spectral resolution
observations are needed to check this result.

The Beppo-SAX data set strong limits on the X-ray reflection
features: EW(Fe-K)$<$15~eV (quasar frame) and $R<0.3$ for the
Compton Hump. 

Hard X-ray observations, as we hoped when initiating this
project, do seem to select physically extreme objects.

%%%%%%%
\acknowledgments

We thank Belinda Wilkes for providing a digital version of her
optical spectrum of the quasar, Daryl Macomb for providing the
EGRET upper limit, Seth Digel for the IRAS data, Gabriele
Ghisselini for the data points in figure~3, and the referee for
a careful reading that improved the paper.
This work was supported in part by NASA contract NAS8-39073 (ASC),
and NASA grants NAGW-2201 (LTSA) and NAGW-  (ADP).
This research has made use of data obtained through the High
Energy Astrophysics Science Archive Research Center Online
Service, provided by the NASA-Goddard Space Flight Center.
This research has made use of the NASA/IPAC Extragalactic
Database (NED) which is operated by the Jet Propulsion
Laboratory, Caltech, under contract with the National Aeronautics
and Space Administration
Based on photographic data obtained using The UK Schmidt
Telescope. The UK Schmidt Telescope was operated by the Royal
Observatory Edinburgh, with funding from the UK Science and
Engineering Research Council, until 1988 June, and thereafter by
the Anglo-Australian Observatory. Original plate material is
copyright (c) the Royal Observatory Edinburgh and the
Anglo-Australian Observatory. The plates were processed into the
present compressed digital form with their permission. The
Digitized Sky Survey was produced at the Space Telescope Science
Institute under US Government grant NAG W-2166.

%%%%%%%

%%%%%%%%%%%%%%
\end{document}